\documentclass[12pt]{article}

\usepackage{scicite}

\usepackage{times}

\usepackage{amsmath,graphicx}
\usepackage{amssymb}  
\usepackage{bm}  
\usepackage[utf8]{inputenc} 

\usepackage{algorithm,algorithmicx,algpseudocode}

\newenvironment{sciabstract}{%
\begin{quote} \bf}
{\end{quote}}

\newcounter{lastnote}

\usepackage{graphicx}		
\graphicspath{{./Figures/}}	
\DeclareGraphicsExtensions{.pdf,.png,.jpg}	
\usepackage[dvipsnames]{color}

\title{HMM for short independent sequences: Multiple sequence Baum-Welch application}

\author
{Margarita Cabrera-Bean,$^{1\ast}$ Josep Vidal,$^{1}$ Sergio Fernández-Bertolín $^{2}$\\ Albert Roso-Llorach$^{2}$, Concepción Violán$^{2,3,4,5}$\\
\\
\normalsize{$^{1}$Dept. of Signal Theory and Communications}\\
\normalsize{ Universitat Polit\`ecnica de Catalunya (UPC), Barcelona, Spain}\\
\normalsize{$^{2}$Institut Universitari d’Investigació en Atenció Primària Jordi Gol,}\\
\normalsize{ IDIAP Jordi Gol, Barcelona, Spain}\\
\normalsize{$^{3}$Germans Trias i Pujol Research Institute (IGTP), Badalona, Spain}\\
\normalsize{$^{4}$Red de Investigacion en Cronicidad, Atención Primaria y }\\
\normalsize{Prevención y Promoción de la Salud (RICCAPPS) (RD21/0016/0029),}\\
\normalsize{Instituto de Salud Carlos III, Madrid, Spain}\\
\normalsize{$^{5}$Departament de Medicina. Universitat Autònoma de Barcelona, Spain}\\
\\
\normalsize{$^\ast$To whom correspondence should be addressed; E-mail:  marga.cabrera@upc.edu.}
}

\date{}

\begin{document} 


\baselineskip24pt


\maketitle


\begin{sciabstract}
  In the classical setting, the training of a Hidden Markov Model (HMM) typically relies on a single, sufficiently long observation sequence that can be regarded as representative of the underlying stochastic process. In this context, the Expectation–Maximization (EM) algorithm is applied in its specialized form for HMMs, namely the Baum–Welch algorithm, which has been extensively employed in applications such as speech recognition. The objective of this work is to present pseudocode formulations for both the training and decoding procedures of HMMs in a different scenario, where the available data consist of multiple independent temporal sequences generated by the same model, each of relatively short duration, i.e., containing only a limited number of samples. Special emphasis is placed on the relevance of this formulation to longitudinal studies in population health, where datasets are naturally structured as collections of short trajectories across individuals with point data at follow-up.
\end{sciabstract}
\section{Introduction}
Hidden Markov Models (HMMs) and their parameter estimation through the Baum–Welch algorithm are well-established probabilistic tools in machine learning, extensively covered in the literature. One of the pioneering works in rigorously describing HMMs and applying them to speech recognition is given in [\citen{Rabiner89}], where Baum-Welch and Viterbi algorithms were updated to solve three basic identified problems for HMMs. In Bishop’s book [\citen{Bishop06}], an entire chapter is dedicated to sequential data, addressing the inference problem in HMMs in detail. Furthermore, a MATLAB implementation of the algorithms presented in the book is openly available on GitHub [\citen{MoChen25}]. The foundational principles of HMMs and the iterative procedure for parameter re-estimation, known as the Baum-Welch algorithm, are also thoroughly discussed in [\citen{Theodoridis2020ML}]. An application example is the study in [\citen{Vio20}], which used HMMs to successfully identify and characterize longitudinal multimorbidity trajectories in patients aged 65–99 years in Catalonia (Spain). Similarly, in [\citen{Roso22}], Hidden Markov Models were applied to model the temporal evolution of both, multimorbidity patterns and individuals' transitions over a 12-year follow-up, working with a Swedish base-data related to Aging and Care in Kungsholmen (SNAC-K) on adults over 60 years.

As highlighted in the abstract, this work generalizes the training procedure of HMMs to scenarios where multiple independent sequences are available, rather than a single long trajectory. Such settings are common in longitudinal population health studies, where individuals are classified into latent health patterns characterized, for instance, by multiple diagnoses and different frailty conditions, and evolve across patterns over time. Typically, each individual contributes a relatively short sequence of length $T$, which is insufficient to represent the full range of model dynamics. However, a large number, $N$ of independent sequences can be collected, each corresponding to a different individual. This motivates the need to extend classical HMM training algorithms, which in the literature are generally formulated for a single long sequence, to handle multiple short, independent sequences effectively.

The paper is structured as follows. Section \ref{sec:SigMod} introduces the signal model that serves as the foundation for the subsequent developments. Section \ref{sec:PbSt} revisits the inference problem and extends the standard probability definitions and recursions to the case of multiple sequences. Section \ref{sec:MSBW} presents the Baum–Welch algorithm generalized to this setting (Multiple Sequence Baum-Welch), while Section \ref{sec:Scaling} incorporates the use of scaling factors to prevent numerical overflow. Section \ref{sec:VA} reviews the Viterbi algorithm for sequence decoding once the model has been estimated. Finally, Section \ref{sec:Prac} discusses practical considerations for implementing the Extended Baum–Welch (EBW) algorithm in a representative application within the domain of population health studies and section \ref{sec:Conc} concludes the work.

\section{Signal Model} \label{sec:SigMod}

A Hidden Markov Model (HMM) is a type of dynamic Bayesian network.
The so-called state observation is modeled by the random variable $\mathbf{y}(t)\in R^D$ where $D$ is the number of scalar (numeric or categorical) observations in each time instant $t$. A latent variable $x(t)\in\{1,..K\}$ is associated with each observation $\mathbf{y}(t)$ where $K$ is the number of states. To clarify, if $x(t)=i$, it means that the hidden state, at time step $t$ is $i$, but we cannot observe it directly, we only can access to the observed vector $\mathbf{y}(t)$. A detailed mathematical expression to indicate an observation vector is

\begin{equation} 
\label{eq:xvector}
\mathbf{y}(t)=\left[ \begin{matrix}
y_1(t)\\
..\\
y_D(t)
\end{matrix}\right]
\end{equation}
where each entry, $y_l(t);l=1,...,D$ is a single observation.

The system dynamics are modelled via the latent variables and observations are considered to be the output of a measuring noisy sensing device.

\section{Inference problem statement} \label{sec:PbSt}
An HMM model is fully described by the following set of parameters and functions:
\begin{enumerate}
	\item Number of states $K$
	\item The probabilities for the initial state $\pi_i\triangleq \Pr\left\{x(t)=i\right\}, i=1,..K$
	\item The set of transitions probabilities $P_{ij}\triangleq\Pr(x(t)=i|x(t-1)=j),\mspace{10mu} \forall i,j=1,..,K$.
	\item The state emission distribution is given by $p\left\{\mathbf{y}(t)|x(t)\right\}$, i.e, the distribution of the observation vector conditioned on the current hidden state $x(t)$. For convenience, when the state is explicitly specified as $x(t)=i$, we write $p_i(\mathbf{y}(t)) \triangleq p\left\{\mathbf{y}(t)|x(t)=i\right\}$ to denote the emission distribution associated with state $i$.
    
\end{enumerate}


Two independent properties usually assumed are
\begin{enumerate}
	\item The future state is independent of the past state given the present state, i.e. $$\Pr\left\{(x(t+1)|x(1),...,x(t) ) \right\}=\Pr\left\{(x(t+1)|x(t)  \right\}$$
	\item The observations are independent of the future and past states given the present state $$p\left\{\mathbf{y}(t)|x(1),...,x(T) \right\}=p\left\{\mathbf{y}(t)|x(t) \right\}$$
\end{enumerate}

The state conditioned observation distribution in the multimorbidity application are modelled as Gaussian distributed, i.e.
\begin{equation} \label{eq:NorDist}
p_i(\mathbf{y}(t))=\mathcal{N}\left(\mathbf{y}(t);\mathbf{m}_i,\mathbf{C}_i\right);\mspace{10mu}i=1,..,K
\end{equation}
where $\mathcal{N}\left(\mathbf{y}(t);\mathbf{m}_i,\mathbf{C}_i\right)$ denotes the well-known Gaussian distribution defined over a $D$-dimensional vector of continuous variables, $\mathbf{y}(t)$ in our case, which is given by
\begin{equation} \label{eq:Gauss}
\mathcal{N}\left(\mathbf{y}(t);\mathbf{m}_i,\mathbf{C}_i\right) \triangleq 
\frac{1}{\sqrt{(2\pi)^D\det(\mathbf{C}_i)}}\exp\left(-\frac{1}{2}(\mathbf{x}-\mathbf{m}_i)^T\mathbf{C}_i^{-1}(\mathbf{x}-\mathbf{m}_i)\right)
\end{equation}
in (\ref{eq:Gauss}) $\mathbf{m}_i$ denotes the mean vector and $\mathbf{C}_i$ the covariance matrix for the observation vector $\mathbf{y}(t)$ associated to the state $x(t)=i;i=1,...,K$.

In the remainder of this document, the notation introduced in equation (\ref{eq:NorDist}) will be used to simplify the intrinsic mathematical formulation as much as possible.

At this point, the model can be represented by the vector $\bm{\theta}$, which encapsulates all the unknown parameters, and is defined as
\begin{equation} \label{eq:Model}
\bm{\theta}\triangleq\{\pi_i,p_{ij},\mathbf{m}_i,\mathbf{C}_i,\forall i,j=1,..,K\}
\end{equation}

Let the observation vector be measured over $T$ discrete time steps across $N$ independent sequences, where each sequence corresponds to a distinct hidden state trajectory. We define $$\mathcal{Y}\triangleq\{\mathbf{y}_n(t),t=1,..,T,n=1,..,N\}$$ as the set of observed vectors, and $$\mathcal{X}\triangleq\{\mathbf{x}_n(t),t=1,..,T,n=1,..,N\}$$ as the set of hidden trajectories.

\paragraph{Problem statement} 

At this stage, the primary objective is twofold: (i) to estimate the set of hidden states $\mathcal{X}$, and (ii) to infer the parameters of the model $\bm{\theta}$.

Below, some variables related to different probabilities, are defined that will be useful later in the development of the algorithms required to address the proposed problem.
The implicit probabilities that arise in the training algorithm development account for the presence of multiple mutually independent sequences, for instance amounting to $N$. Consequently, they represent a generalization of the plain case of a single sequence ($N=1$), which is the scenario most commonly found in the literature [\citen{Duda2001}].

\paragraph{Join probability} 

The joint probability of the set of observations and the set of trajectories is
\begin{equation} \label{eq:JoinP}
p\left\{\mathcal{Y,X}\right\}=
\prod_{n=1}^N\Pr\left\{x_n(1)\right\}p\left\{\mathbf{y}_n(1)|x_n(1)\right\}
\prod_{t=2}^{T}\Pr\left\{x_n(t)|x_n(t-1)\right\}p\left\{\mathbf{y}_n(t)|x_n(t)\right\}
\end{equation}
where, regarding a single sequence $n^{th}$, composed by $\{\mathbf{y}_n(t),t=1,..,T\}$ the state emission distribution at time step $t$ given the current hidden state $x_n(t)$ is given as
\begin{equation}
p\left\{\mathbf{y}_n(t)|\mathbf{x}_n(t)\right\}=
\prod_{i=1}^{K}\left(\mathcal{N}\left(\mathbf{y}_n(t);\mathbf{m}_i,\mathbf{C}_i\right)\right)^{z_i(x_n(t))}=
\prod_{i=1}^{K}p_i(\mathbf{y}_n(t))^{z_i(x_n(t))}
\end{equation}
Here, function $z_i(x_n(t))$ equals 1 when $x_n(t)=i$ and 0 otherwise.
\paragraph{Forward Iteration, Multiple Sequence}

A forward variable is defined as the probability of partial observation of sequence $\mathbf{y}_n(t)$ until time $t$ and state $x_n(t)=i$, given the model represented by the parameter set $\bm{\theta}$ defined in (\ref{eq:Model}):
\begin{equation} \label{eq:alpha}
\alpha_n\left(t,i\right)\triangleq
p\left\{\mathbf{y}_n(1),..,\mathbf{y}_n(t),x^n(t)=i|\bm{\theta}\right\}
=p_i\left(\mathbf{y}_n(t)\right)\sum_{j=1}^{K}\alpha_n\left(t-1,j\right)P_{ij};\mspace{10mu} t=2,..T
\end{equation}
with 
\begin{equation*}
\alpha_n\left(1,i\right)\triangleq p\left\{\mathbf{y}_n(1),x_n(1)=i|\bm{\theta}\right\}\pi_i
\end{equation*}
and
\begin{equation*}
p\left\{\mathbf{y}_n(1),..,\mathbf{y}_n(T)|\bm{\theta}\right\}=
\sum_{i=1}^{K}\alpha_n\left(T,i\right)
\end{equation*}

\paragraph{Backward Iteration, Multiple Sequence}

Similarly, a backward variable is defined as the probability of partial observation of sequence $\mathbf{y}_n(t)$ from time $t+1$ to the end $T$, given the state $x_n(t)=i$ and the model $\bm{\theta}$

\begin{equation} \label{eq:beta}
\begin{gathered}
\beta_n\left(t,i\right)\triangleq p\left\{\mathbf{y}_n(t+1),..,\mathbf{y}_n(T),x_n(t)=i|\bm{\theta}\right\}\\
=\sum_{j=1}^{K}\beta_n\left(t+1,j\right)P_{ji}p_j\left(\mathbf{y}_n(t+1)\right);\mspace{10mu} t=T-1,..,1
\end{gathered}
\end{equation}
with
\begin{equation*}
\beta_n\left(T,i\right)\triangleq 1;\mspace{10mu}i=1,..K
\end{equation*}

\paragraph{Smoothing Recursion, Multiple Sequence}
Combining both, the forward and the backward variables, we obtain the probability of state i at time step t given the observations $\mathbf{y}_n(t)$ and the model $\bm{\theta}$.

\begin{equation} \label{eq:gamma}
\gamma_n\left(t,i\right)\triangleq \Pr\left\{x_n(t)=i|\mathbf{y}_n(1),..,\mathbf{y}_n(T);\bm{\theta}\right\}
=\frac{\alpha_n\left(t,i\right)\beta_n\left(t,i\right)}{\sum_{j=1}^{K}\alpha_n\left(t,j\right)\beta_n\left(t,j\right)}
\end{equation}
Note that $\sum_{i=1}^{K}\gamma_n(t,i)=1$ for $t=1,..,T$

Finally, the probability of the system being at states $j$ and $i$ at time steps $t-1$ and $t$, respectively, conditioned on the transmitted sequence of observations, $\mathbf{y}_n(t)$, is given by

\begin{equation}\label{eq:xi}
\begin{gathered}
\xi_n\left(t-1,t,j,i\right)\triangleq
\Pr\left\{x_n(t-1)=j,x_n(t)=i|\mathbf{y}_n(1),..,\mathbf{y}_n(T);\bm{\theta}\right\}=\\
=\frac{\alpha_n\left(t-1,j\right)P_{ij}p_i\left(\mathbf{y}_n(t)\right)\beta_n\left(t,i\right)}
{p\left\{\mathbf{y}_n(1),..,\mathbf{y}_n(T)\right\}}=\\
=\frac{\alpha_n\left(t-1,j\right)P_{ij}p_i\left(\mathbf{y}_n(t)\right)\beta_n\left(t,i\right)}
{\sum_{k=1}^K\sum_{l=1}^K\alpha_n\left(t-1,k\right)P_{kl}p_l\left(\mathbf{y}_n(t)\right)\beta_n\left(t,l\right)}
\end{gathered}
\end{equation}

\section{Multiple Sequence Baum-Welch algorithm} \label{sec:MSBW}

In this section, we present the theoretical optimal solution for estimating the model $\bm{\theta}$, from which the Baum–Welch algorithm is derived, i.e., the specific application of the Expectation–Maximization (EM) algorithm designed HMMs. This algorithm is widely used in practice, as it provides a practical approach to the problem, given the impossibility of obtaining a closed-form theoretical solution.

Since this work extends the Baum–Welch algorithm to the case of multiple independent sequences, we will refer to it as the Multiple Sequence Baum–Welch (MSBW) algorithm, to distinguish it from the standard version commonly used.

The objective function to be maximised is the likelihood of the joint probability distribution of observations and hidden states conditioned by the model, that can be derived as the natural logarithm of the probability $p\left\{\mathcal{Y,X}|\bm{\theta}\right\}$. Defining latent variables $z_i(x_n(t))$ as 1 if $x_n(t)=i$ and 0 otherwise, they become indicator (or one-hot) random variables representing the hidden states. Then, the log-likelihood function is defined as

\begin{equation} \label{eq:LogLik}
\begin{gathered}
\log p\left\{\mathcal{Y,X}|\bm{\theta}\right\}=\\
=\text{ln}\prod_{n=1}^{N}\prod_{i=1}^{K}\left(\pi_ip_i\left(\mathbf{y}_n(t)\right)\right)^{z_i(x_n(1))}
\prod_{t=2}^{T}\prod_{i=1}^{K}\prod_{j=1}^{K}P_{ij}^{z_j(x_n(t-1))z_i(x_n(t))}
\prod_{t=2}^{T}\prod_{i=1}^{K}\left(p_i\left(\mathbf{y}_n(t)\right)\right)^{z_i(x_n(t))}=\\
=\sum_{n=1}^{N}\sum_{i=1}^{K}z_i(x_n(1))\left(\log\pi_i+\log p_i\left(\mathbf{y}_n(t)\right)\right)+\\
\sum_{n=1}^{N}\sum_{t=2}^{T}\sum_{i=1}^{K}\sum_{j=1}^{K}z_j(x_n(t-1))z_i(x_n(t))\log P_{ij}+
\sum_{n=1}^{N}\sum_{t=2}^{T}\sum_{i=1}^{K}x_i^n(t)\log p_i\left(\mathbf{y}_n(t)\right)
\end{gathered}
\end{equation}
Note that (\ref{eq:LogLik}) is a detailed expansion of the natural logarithm of the term given in equation \eqref{eq:JoinP}, where the dependence on the model vector $\bm{\theta}$, has been explicitly expressed as a condition in the probability.

Ideally, the model vector should correspond to the value that maximizes the likelihood function \eqref{eq:LogLik}. However, this is computationally infeasible. Since a closed-form solution for the parameters does not exist, the Expectation-Maximization algorithm (EM) is applied to obtain an estimate of the model parameters. The EM is a general approach for finding maximum likelihood (ML) or maximum a posteriori (MAP) estimates of model parameters when the data are incomplete, contain hidden variables, or have latent structures. The key idea is to iteratively improve estimates of the parameters by alternating between computing expected sufficient statistics (E-step) and maximizing the likelihood with respect to the parameters (M-step).

\paragraph{E Step}
Given the current estimate of the parameters $\bm{\theta}^l$, the expected value of the complete-data log-likelihood, conditioned on the observed data and the current parameter estimates is computed. Intuitively, in this step the algorithm "fills in" or estimates the hidden states using the current model.

Mathematically, the function computed is the expectation $\mathcal{Q}(\bm{\theta},\bm{\theta}^{l})\triangleq\mathbb{E}[\log p\left\{\mathcal{Y,X}|\bm{\theta}\right\}]$ with respect to $\Pr\left\{\mathcal{X}|\mathcal{Y};\bm{\theta}^{l}\right\}$.
\begin{equation}
\label{eq:fQ}
\begin{gathered}
\mathcal{Q}(\bm{\theta},\bm{\theta}^{l})=
\sum_{n=1}^{N}\sum_{i=1}^{K}\mathbb{E}\left[z_i(x_n(1))\right]\left(\log\pi_i+\log p_i\left(\mathbf{y}_n(1)\right)\right)+\\
\sum_{n=1}^{N}\sum_{t=2}^{T}\sum_{i=1}^{K}\sum_{j=1}^{K}\mathbb{E}\left[z_j(x_n(t-1))z_i(x_n(t))\right]\log P_{ij}+
\sum_{n=1}^{N}\sum_{t=2}^{T}\sum_{i=1}^{K}\mathbb{E}\left[z_i(x_n(t))\right]\log p_i\left(\mathbf{y}_n(t)\right)
\end{gathered}
\end{equation}
In fact, this step basically consists in computing the expectations,

\begin{equation}\label{eq:EStep}
\begin{gathered}
\mathbb{E}\left[z_i(x_n(t))\right]=\gamma_n\left(t,i;\bm{\theta}^l\right)\mspace{10mu}i=1,..K\\
\mathbb{E}\left[z_j(x_n(t-1))z_i(x_n(t))\right]=
\xi_n\left(t-1,t,j,i;\bm{\theta}^l\right)\mspace{10mu}i,j=1,..K
\end{gathered}
\end{equation}
where the unobservable binary counts of states/transitions (latent variables) are replaced with their expected values ($\gamma_n\left(t,i;\bm{\theta}^l\right)$ and $\xi_n\left(t-1,t,j,i;\bm{\theta}^l\right)$ probabilities), calculated using the Forward-Backward algorithm,  \eqref{eq:gamma} and \eqref{eq:xi}. This allows the subsequent M-step to treat these expectations as fractional counts to re-estimate the model parameters $\bm{\theta}$. Additionally, in \eqref{eq:EStep} we have explicately included the dependency of $\gamma_n\left(t,i;\bm{\theta}^l\right)$ and $\xi_n\left(t-1,t,j,i;\bm{\theta}^l\right)$, defined in \eqref{eq:gamma} and \eqref{eq:xi}, with respect to the current model estimate $\bm{\theta}^l$.

\paragraph{M Step}
At the M-step the gradient of (\ref{eq:fQ}) with regard the parameters is equated to zero. Intuitively, the algorithm updates the parameters to best explain both the observed vectors and the "filled-in" hidden states from the E-step.In our case, the resulting estimated parameters are:
\begin{equation}\label{eq:grad}
\begin{gathered}
\hat{\pi}_i^{l+1}=\frac{\sum_{n=1}^{N}\gamma_n\left(1,i;\bm{\theta}^l\right)}
{\sum_{n=1}^{N}\sum_{j=1}^{K}\gamma_n\left(1,j;\bm{\theta}^l\right)}\mspace{10mu}i=1,..,K\\
\hat{P}_{ij}^{l+1}=\frac{\sum_{n=1}^{N}\sum_{t=2}^{T}\xi_n\left(t-1,t,j,i;\bm{\theta}^l\right)}
{\sum_{n=1}^{N}\sum_{t=2}^{T}\sum_{k=1}^{K}\xi_n\left(t-1,t,j,k;\bm{\theta}^l\right)}\mspace{10mu}i,j=1,..,K\\
\hat{\mathbf{m}}_i^{l+1}=\frac{\sum_{n=1}^{N}\sum_{t=1}^{T}\gamma_n\left(t,i;\bm{\theta}^l\right)\mathbf{y}_n(t)}
{\sum_{n=1}^{N}\sum_{t=1}^{T}\gamma_n\left(t,i;\bm{\theta}^l\right)}\mspace{10mu}i=1,..,K
\\
\hat{\mathbf{C}}_i^{l+1}=\frac{\sum_{n=1}^{N}\sum_{t=1}^{T}\gamma_n\left(t,i;\bm{\theta}^l\right)
	\left(\mathbf{y}_n(t)-\hat{\mathbf{m}}_k^{l}\right)\left(\mathbf{y}_n(t)-\hat{\mathbf{m}}_k^{l}\right)^T}
{\sum_{n=1}^{N}\sum_{t=1}^{T}\gamma_n\left(t,i;\bm{\theta}^l\right)}\mspace{10mu}i=1,..,K
\end{gathered}
\end{equation}

In summary, training an HMM, by applying the Baum-Welch algorithm, comprises the following steps

\begin{enumerate}
	\item Initialize $\bm{\theta}^0$; $l=1$
	\item Compute $\gamma\left(t,i;\theta^{l-1}\right)$ and $\xi\left(t-1,t,j,i;\theta^{l-1}\right)\mspace{10mu}i,j=1,..,K$ by applying \eqref{eq:gamma} and \eqref{eq:xi} 
	\item Update $\bm{\theta}^{l}$ by applying \eqref{eq:grad}
	\item $l \gets l+1$
	\item Repeat 2,3,4 until convergence (i.e., the increase in the log-likelihood function is below a threshold) or a maximum number of iterations is reached $l=iterMAX$.
\end{enumerate}

\section{Multiple Sequence Baum-Welch algorithm using scaling factors} \label{sec:Scaling}
In the Baum–Welch algorithm, scaling factors are introduced to address numerical underflow when computing the forward and backward variables over long observation sequences. Since transition and emission probabilities are often very small, their repeated multiplication can drive values to zero, making the algorithm unstable. To prevent this, at each time step the variables are normalized by a scaling factor, typically the sum of the forward variables. This rescaling keeps values within a manageable range while preserving the correct probability ratios. As a result, the computed expectations and parameter updates remain accurate, with scaling cancelling out in the final calculations. The application of scaling factors to the MSBW algorithm is presented below.

The scaling factors $c_n(t)\triangleq p \left\{\mathbf{y}_n(t)|\mathbf{y}_n(1),..,\mathbf{y}_n(t-1)\right\}$ are defined as the conditional distributions over the observed variables. Note that  $p \left\{\mathbf{y}_n(1),..,\mathbf{y}_n(t)\right\}=\prod_{t'=1}^{t}c_n(t')$ and $p \left\{\mathcal{Y}\right\}=\prod_{n=1}^{N}\prod_{t=1}^{T}c_n(t)$.

The scaled forward variables ($\hat{\alpha}_n(t,i)$) are included in the forward recursion as

\begin{equation}
\begin{gathered}
\alpha_n(t,i)=\hat{\alpha}_n(t,i)\prod_{t'=1}^{t}c_n(t')\\
c_n(t)\hat{\alpha}_n(t,i)=p_i\left(\mathbf{y}(t)\right)\sum_{j=1}^{K}\hat{\alpha}_n\left(t-1,j\right)P_{ij}
\end{gathered}
\end{equation}
and the scaled backward variables $\hat{\beta}_n(t,i)$, as
\begin{equation}
\begin{gathered}
\beta_n(t,i)=\hat{\beta}_n(t,i)\prod_{t'=t+1}^{T}c(t')\\
c_n(t+1)\hat{\beta}_n(t,i)=\sum_{j=1}^{K}\hat{\beta}_n\left(t+1,j\right)P_{ji}p_j\left(\mathbf{y}_n(t+1)\right)
\end{gathered}
\end{equation}
The required marginal variables are given by
\begin{equation}
\begin{gathered}
\gamma_n\left(t,i\right)=\hat{\alpha}_n(t,i)\hat{\beta}_n(t,i)\\
\xi_n\left(t-1,t,j,i\right)=\frac{1}{c_n(t)}\hat{\alpha}_n\left(t-1,j\right)P_{ij}p_i\left(\mathbf{y}_n(t)\right)\hat{\beta}_n\left(t,i\right)
\end{gathered}
\end{equation}

The Baum–Welch algorithm with scaling factors is presented in algorithm \ref{alg:BWalg}. For the sake of clarity and to facilitate implementation, the table uses the notation $\mathcal{N}(\mathbf{y}_n(t);\mathbf{m}_i^{l-1},\mathbf{C}_i^{l-1})$ to denote the state-conditioned observation distribution, rather than the shorthand $p_i\left(\mathbf{y}_n(t)\right)$ defined in \eqref{eq:NorDist}. This choice makes explicit the dependence on the current estimates of the mean and covariance matrices, $\mathbf{m}_i^{l-1},\mathbf{C}_i^{l-1}$ within the conditioned distributions.

\begin{algorithm}
	\caption{Multiple Sequence Baum Welch}\label{alg:BWalg}
	\begin{algorithmic}[1]
		 \Require $\mathcal{Y},\bm{\theta}^0$
		 \Ensure $\mathcal{Q}(\bm{\theta},\bm{\theta}^{l}),\bm{\theta}^{iterMAX}$
		\Procedure{BW}{}
		\State $llh(1,..,iterMAX)=-\inf;$ \Comment{Likelihood Initialization}
		\State $\epsilon = 1e-4; iterMAX=1000; iterMIN=10$;
		\State $l \gets 0$
		\Repeat
		\State $l \leftarrow l+1$
		\State \textit{Compute} $\mathcal{N}(\mathbf{y}_n(t);\mathbf{m}_i^{l-1},\mathbf{C}_i^{l-1});\forall n=1,..N;t=1,..T,i=1,..K$\Comment{E-Step}
		
		\For{$n=1,..,N$}
		\State $\alpha_n(1,i)=\pi^{l-1}_i\mathcal{N}(\mathbf{y}_n(1);\mathbf{m}_i^{l-1},\mathbf{C}_i^{l-1});\forall i=1,..K$\Comment{Forward Variable}
		\State $c_n(1)=\sum_{i=1}^{K}\alpha_n(1,i)$
		\State $\alpha_n(1,i)=\frac{\alpha_n(1,i)}{c_n(1)};\forall i=1,..K$
		\For{$t=2,..,T$}
		\State $\alpha_n(t,i)=\mathcal{N}(\mathbf{y}_n(t);\mathbf{m}_i^{l-1},\mathbf{C}_i^{l-1})\sum_{j=1}^{K}\alpha_n\left(t-1,j\right)P_{ij};\forall i=1,..K$
		\State $c_n(t)=\sum_{i=1}^{K}\alpha_n(t,i)$
		\State $\alpha_n(t,i)=\frac{\alpha_n(t,i)}{c_n(t)};\forall i=1,..K$
		\EndFor
		
		\State $\beta_n(T,i)=1;\forall i=1,..K$\Comment{Backward Variable}
		\For{$t=T-1,-1:1$}
		\State $\beta_n(t,i)=\frac{\sum_{j=1}^{K}\beta_n(t+1,j)P_{ji}\mathcal{N}(\mathbf{y}_n(t+1);\mathbf{m}_j^{l-1},\mathbf{C}_j^{l-1})}{c_n(t+1)};\forall i=1,..K$
		\EndFor

		\For{$t=1,..,T$}\Comment{Smoothing Recursion}
		\State $\gamma_n(t,i)=\alpha_n(t,i)\beta_n(t,i);\forall i=1,..K$
		\EndFor
		
		\EndFor
		
		\State $llh(l)=\frac{1}{N}\frac{1}{T}\sum_{n=1}^{N}\sum_{t=1}^{T}\text{ln} c_n(t)$\Comment{Likelihood function and check convergence}
		\If {$l>=iterMIN$ and $llh(l)-llh(l-1)<\epsilon|llh(l-1)|$}
			\State  $l \gets iterMAX$
		\EndIf
		
		\State $\pi_i^l=\frac{1}{N}\sum_{n=1}^{N}\gamma_n(1,i)$ \Comment{M-Step}
		\State $P_{ij}^l=\sum_{n=1}^{N}\sum_{t=1}^{T-1}
		\frac{\alpha_n(t,j)P_{ij}^{l-1}\mathcal{N}(\mathbf{y}_n(t+1);\mathbf{m}_i^{l-1},\mathbf{C}_i^{l-1})\beta_n(t+1,i)}{c_n(t+1)};\forall i,j=1,..K$
		\State $P_{ij}^l=\frac{P_{ij}^l}{\sum_{m=1}^{K}P_{mj}^l};\forall  j=1,..K$
		\State $\mathbf{m}^l_i=\frac{\sum_{n=1}^{N}\sum_{t=1}^{T}\gamma_n(t,i)\mathbf{y}_n(t)}
		{\sum_{n=1}^{N}\sum_{t=1}^{T}\gamma_n(t,i)};\forall  i=1,..K$
		\State $\mathbf{C}^l_i=\frac{\sum_{n=1}^{N}\sum_{t=1}^{T}\gamma_n(t,i)(\mathbf{y}_n(t)-\mathbf{m}^l_i)(\mathbf{y}_n(t)-\mathbf{m}^l_i)^T}
		{\sum_{n=1}^{N}\sum_{t=1}^{T}\gamma_n(t,i)};\forall  i=1,..K$
		\Until $l=iterMAX$
		\EndProcedure
	\end{algorithmic}
\end{algorithm}

\section{Viterbi Algorithm to estimate hidden states} \label{sec:VA}
Once the model $\bm{\theta}$ has been estimated by applying the Baum-Welch algorithm, it is used to estimate the set of hidden states trajectories $\mathcal{X}$ from the observed data set $\mathcal{Y}$.
To do it, the Viterbi algorithm (VA) can be used to find the most probable sequence of states. So, the current goal is to find each ${x_n(1),..,x_n(T)}$ given the observed sequence ${\mathbf{y}_n(1),..,\mathbf{y}_n(T)}$ and the model $\bm{\theta}$ for $n=1,...,N$. The best score or highest probability along a single path at time $t$, which ends in state $i$ is defined as $\delta_n(t,i)$

\begin{equation}
\delta_n(t,i)=\max_{x_n(1),..,x_n(t-1)} \Pr\left\{x_n(1),..,x_n(t-1),x_n(t)=i|\mathbf{y}_n(1),..,\mathbf{y}_n(t);\bm{\theta}\right\}
\end{equation}
This function can be computed by induction using the VA algorithm, which main steps are described below:

\paragraph{Initialization}
\begin{equation}
\delta_n(1,i)=\pi_i p\left\{\bm{y}_n(1)|x_n(1)=i;\bm{\theta}\right\}\mspace{10mu}i=1,..K
\end{equation}
To keep track of the path the variable $\psi_n(t,i)$ is started as
\begin{equation*}
\psi_n(1,i)=0\mspace{10mu}i=1,..K
\end{equation*}
\paragraph{Recursion}
\begin{equation}
\begin{gathered}
\delta_n(t,i)=\max_{j}\left[\delta_n(t-1,j)P_{ij}\right]
p\left\{\mathbf{y}_n(t)|x_n(t)=i;\bm{\theta}\right\}\mspace{10mu}i=1,..K,t=2,..T;\\
\psi_n(t,i)=\arg \max_{j}\left[\delta_n(t-1,j)P_{ij}\right]\mspace{10mu}i=1,..K
\end{gathered}
\end{equation}
\paragraph{Termination and backtracking}
\begin{equation}
\begin{gathered}
\hat{x}_n(T)=\arg \max_{j}\left[\delta_n(T,j)\right]\\
\hat{x}_n(t)=\psi_n(t+1,\hat{x}_n(t+1))\mspace{10mu}t=T-1:-1:1
\end{gathered}
\end{equation}
Algorithm \ref{alg:Valg} presents a detailed pseudocode of the Viterbi algorithm. For clarity, the subindex $n$ has been omitted from the notation, as the $N$ observation sequences are processed independently. Accordingly, the algorithm is executed iteratively, once for each observation sequence.
\begin{algorithm}
	\caption{Viterbi Algorithm}\label{alg:Valg}
	\begin{algorithmic}[1]
		\Require $\mathcal{Y}=\{\mathbf{y}(t),t=1,..,T\},\bm{\theta}$
		\Ensure $\hat{\mathcal{X}}=\{\hat{\mathbf{x}}(t),t=1,..,T\},$
		\Procedure{VA}{}
		\For{$i=1,..,K$} \Comment{Initialization}
		\State $\delta(1,i)=\log \left(\pi_i
		\mathcal{N}(\mathbf{y}(1);\mathbf{m}_i,\mathbf{C}_i)\right)$
		\State $\psi(1,i)=0$
		\EndFor
		\For{$t=2,..,T$} \Comment{Recursion}
		\For{$i=1,..,K$}
		\State $\delta(t,i)=\max_{j}\left[\delta(t-1,j)+\log P_{ij}\right]
		+\log\left(\mathcal{N}(\mathbf{y}(t);\mathbf{m}_i,\mathbf{C}_i)\right)$
		
		\State $\psi(t,i)=\arg \max_{j}\left[\delta(t-1,j)+\log P_{ij}\right]$
		\EndFor
		\EndFor
		\State $\hat{x}(T)=\arg \max_{j}\left[\delta(T,j)\right]$\Comment{Termination and Backtracking}
		\For{$t=T-1:-1:1$} 
		\State $\hat{x}(t)=\psi(t+1,\hat{x}(t+1))$
		\EndFor
		\EndProcedure
	\end{algorithmic}
\end{algorithm}

\section{Implementation Details and Practical Considerations} \label{sec:Prac}
While the previous sections focused on the theoretical formulation of the involved algorithms, their practical deployment requires addressing several implementation details. Most of considerations included in this section complement the explained algorithm in real-world applications. The first present issue is the applicability of the HMMs in a generic and typical health analysis and the rest of them are devoted to some useful details.

\paragraph{A practical use in population health monitoring}
The health of a population of size $N$ individuals is to be monitored over a period of $T$ years. To model this, the population is classified into $K$ health patterns, each representing a distinct state. Specifically, if individual $n$ in year $t$ is in pattern $i$ with $i\in{1,…,K}$, this is represented by the variable $x_n(t)=i$. Collectively, the state variables for all individuals form $N$ sequences of length $T$, which are grouped into the set $\mathcal{X}$. Since the true state or pattern of each individual is not directly observable, $\mathcal{X}$ is treated as the set of latent variables, or hidden states. Instead, for each individual and year, a set of $D$ observable variables is available, grouped into the feature vector $\mathbf{y}_n(t)$ These variables may include, for instance, the presence or absence of specific diseases, the use of certain medications, the results of clinical tests, the number of specialist visits for a particular condition, or the number of hospitalizations and clinical events such as heart attacks or strokes. Thus, the observation vectors provide a data-driven characterization of health status throughout the study period. All such observation vectors are grouped into the set $\mathcal{Y}$. $\mathcal{X}$ and $\mathcal{Y}$ denote, respectively, the set of hidden state trajectories and the set of observation vectors introduced in Section \ref{sec:PbSt}.

\paragraph{Modeling an absorbing state}

In the context of Hidden Markov Models (HMMs), a state is defined as absorbing if, once entered, in cannot be left; formally, the self-transition probability satisfies $P_{i,i}=1$ and $P_{j,i}=0$ if $i\neq j$. In other words, in an HMM an absorbing state represents a condition or situation in the modeled system that, once reached, the process remains there forever, regardless of future time steps. For instance, if in a health monitoring HMM, there is included a "death" state, it would typically be modeled as absorbing, since once a person is dead, enters it, and no transitions to other health states occur. Let's assume that the state $K^{th}$ is an absorbing one. Certain parameters in the model's parameter set are fixed as shown in \eqref{eq:IniAbs} and, as a result, they are no longer included in the model vector $\bm{\theta}$ defined in \eqref{eq:Model}.

\begin{equation}
\label{eq:IniAbs}
\begin{gathered}
\pi_K=0\\
P_{iK}=0\mspace{10mu}i=1,..,K-1\\
P_{KK}=1
\end{gathered}
\end{equation}

The first condition in (\ref{eq:IniAbs}) assumes that at time $t=0$ there are not dead people in the population and the other conditions denote that once a person has arrived at the $K^{th}$ state he or she remains in it forever. Indeed, we do not need to compute $\mathbf{m}^l_K$ and $\mathbf{C}^l_K$. Since there are not available data for dead people, once a person is dead, a zero vector will be assigned to the corresponding observation vectors. In other words, if a person enumerated as $n$ dies at year $t_m$, then $\mathbf{y}_n(t)=\mathbf{0}, t_m<t\leq T$, where $\mathbf{0}\in R^D$ denotes a vector whose entries are all zero. Finally, in order to maintain the pseudocode given in algorithm \ref{alg:BWalg} the state emission distributions are redefined considering the absorbing state $K$ as

\begin{equation}
\label{eq:GenGauss}
\begin{gathered}
p_i(\mathbf{y}_n(t))=\mathcal{N}(\mathbf{y}_n(t);\mathbf{m}_i,\mathbf{C}_i)
\mspace{10mu}\text{if}\mspace{10mu}\mathbf{y}_n(t)\neq \mathbf{0}\mspace{10mu}i=1,..,K-1\\
p_i(\mathbf{y}_n(t))=0
\mspace{10mu}\text{if}\mspace{10mu}\mathbf{y}_n(t)= \mathbf{0}\mspace{10mu}i=1,..,K-1\\
p_K(\mathbf{y}_n(t))=0\mspace{10mu}\text{if}\mspace{10mu}\mathbf{y}_n(t)\neq \mathbf{0}\\
p_K(\mathbf{y}_n(t))=1\mspace{10mu}\text{if}\mspace{10mu}\mathbf{y}_n(t)= \mathbf{0}
\end{gathered}
\end{equation}

\paragraph{Covariance Matrix Simplification}

This simplification considerably reduces the number of parameters included in the model vector, although it may impact the accuracy of the parameters estimated from the actual model. Line 33 in algorithm \ref{alg:BWalg} would be modified as

\begin{equation}
\begin{gathered}
\mathbf{C}^l_i=\text{diag}\left(\sigma ^l_i(1),..,\sigma ^l_i(D)\right)\forall  i=1,..K\\
\sigma ^l_i(d)=\frac{\sum_{n=1}^{N}\sum_{t=1}^{T}\gamma_n(t,i)(y_{n_d}(t)-m^l_i(d))^2}
{\sum_{n=1}^{N}\sum_{t=1}^{T}\gamma_n(t,i)}
\end{gathered}
\end{equation}

\paragraph{Model validation}
The supplementary information of [\citen{Vio20}] describes a validation methodology used to assess the results of Algorithm \ref{alg:BWalg}. This approach compares the algorithm's output on real data against a synthetic model that contains deliberate misadjustment when calculating the likelihood function $ p\left\{\mathcal{Y,X}|\bm{\theta}\right\}$. This measurement effectively quantifies how well the computed model fits the observed real data.

\section{Conclusions} \label{sec:Conc}
In conclusion, this document has presented the methodology for learning a Hidden Markov Model when the available training data consist of multiple mutually independent sequences, all generated by the same underlying model. The resulting algorithm has been referred to here as the Multiple Sequence Baum–Welch (MSBW) to distinguish it from its plain version. This algorithm is particularly well suited to common scenarios in healthcare applications, where patient databases are used to monitor disease progresion  in individuals  through different disease patterns. The data obtained from each patient forms a sequence independent of the others.

\bibliographystyle{IEEEtran}
\bibliography{scibib}

\end{document}